# Engineering liquid-liquid interfaces for high-entropy alloy synthesis

**Authors:**

Qiubo Zhang[1], Yi Chen[1,2], Karen. C. Bustillo[3], and Haimei Zheng[1,2]*

**Affiliations:**

[1]Materials Science Division, Lawrence Berkeley National Laboratory, Berkeley, CA 94720, USA.

[2]Department of Materials Science and Engineering, University of California, Berkeley, CA 94720, USA.

[3]National Center for Electron Microscopy, Molecular Foundry, Lawrence Berkeley National Laboratory, Berkeley, CA 94720, USA.

*Corresponding author. Email: hmzheng@lbl.gov

**Abstract:**

High-entropy alloys (HEAs) with various promising applications have attracted significant interest. However, alloying immiscible metal elements while controlling the morphology and crystallinity remains extremely challenging. We report a general route, by engineering liquid-liquid interfaces, for the synthesis of HEAs with controlled crystallinity (single crystal, mesocrystal, polycrystal, amorphous), morphology (0-dimension, 2-dimension, 3-dimension), and high composition diversity (20 elements) under mild conditions (room temperature to 80 °C). As reactions can be spatially confined at the liquid-liquid interface, it provides an opportunity to control the kinetics and reduce the alloying temperature. Our real-time observation of the alloying of HEAs reveals hydrogen-enhanced mixing and isothermal solidification that kinetically traps the mixing states due to composition changes. The concept of liquid-liquid interface engineering can be applied to obtain other high-entropy compounds.

**One-Sentence Summary:** We discovered a general route for the synthesis of HEAs with controlled crystallinity and morphology under mild conditions.

**Main Text:**

High-entropy alloy (HEA) nanomaterials, with more than five metal elements mixed together, have potential applications ranging from catalysis (*1-6*) to batteries (*7*) and others (*8*). Due to the thermodynamic immiscibility of certain elements, non-equilibrium methods have been developed to kinetically trap the high-entropy states (*2, 9*). For example, effective mixing of metal elements can be achieved at high temperatures, followed by subsequent fast cooling such that the high-entropy phase can be trapped at room temperature (*2, 10-12*). The liquid metal-assisted strategy facilitates the synthesis of HEA nanoparticles (HEA-NPs) at a lower temperature (e.g., 923 K) (*9*). However, HEAs often retain the spherical shape inherited from their states at melting, exhibiting either a single-crystalline or amorphous structure (*2, 9-12*). The limited morphological and structural variations, as well as the lack of crystallinity control, restrict their applications. Wet-chemistry approaches provide versatility of particle sizes, morphologies, and structures at low temperatures (*6, 13-15*). However, wet-chemistry methodologies can only be applied to specific systems, proving unsuitable for immiscible elemental combinations. To date, developing a general strategy that can precisely control the elemental composition, morphology, and crystallinity of HEAs under mild conditions is intriguing as well as challenging.

Here, we introduce our liquid-liquid interface engineering approach, which takes advantage of liquid metal (e.g., gallium (Ga) or Ga-based alloys; **fig. S1-3**) and the unique liquid-liquid interfaces between the liquid metal and a non-metallic solvent. Ga or Ga-based alloy liquid metals are dense and interact strongly with other elements, thus serving as ideal solvents for the dynamic mixing of solute metal elements (*16, 17*). By utilizing the liquid-liquid interfacial reactions (e.g., Ga or other reducing agents can reduce metal ions in aqueous solution), the reduction of metal precursors can be restricted to two-dimensional (2D) interfaces between liquid metal and non-metallic liquid solvents. The product metal atoms dissolve into the metallic liquid solvent, refreshing the reaction continuously. Consequently, the alloying process can be achieved at low temperatures (e.g., room temperature). The reduction reaction includes the consumption of liquid metal, the integration of product metal atoms, and the generation of hydrogen gases ($H_2$), thereby altering the metal composition and inducing a shift in the melting point of the liquid alloy. Thus, it enables the isothermal solidification to kinetically trap the high entropy states. By engineering the liquid-liquid interface reaction kinetics using coordinated metal reduction, diffusion of atoms in the liquid metal solvent, and isothermal solidification, emergent HEA nanostructures with designed crystallinity and morphology can be achieved.

**The liquid-liquid interface engineering method**

We developed a liquid-liquid interface engineering method that employs site-specific reduction and isothermal solidification of metal precursors in a liquid metal solvent to produce HEA nanomaterials with distinct morphologies and crystallinity under mild conditions (**Fig. 1**). As shown in **Fig. 1A**, we first load the liquid metal (Ga or Ga-based alloy) onto a carbon substrate. Then, we heat the substrate and stabilize it at mild conditions (e.g., room temperature to 80 °C) for two minutes. Subsequently, an aqueous salt solution containing various metal precursors ($MCl_xH_y$, where M represents platinum (Pt), gold (Au), copper (Cu), lead (Pb), and others) was dropped onto the substrate, thereby enveloping the liquid metal and giving rise to 2D interfaces between the liquid metal and the aqueous solution. The metal ions in the aqueous solution can be reduced at the interface to generate metal atoms, which are incorporated into the Ga-based liquid metal spontaneously due to the ability of liquid metal to dissolve various metal atoms (*17, 18*). As the liquid metal is continuously consumed and metal atoms constantly infiltrate, the melting point of the liquid alloy increases (*19, 20*). Once the melting point exceeds the sample temperature, the liquid alloy with a high entropy state stabilizes into a high-entropy alloy (HEA).

Since the reduction reactions of the metal ions only occur at the liquid-liquid interface (2D) rather than in the bulk liquid (3D), it drastically reduces the temperature and time required for mixing different metals as those in bulk (3D) reactions. The reaction is described as:

$$Ga + \frac{3}{x}M^{x+} \rightarrow Ga^{3+} + \frac{3}{x}M \quad (1)$$

, where $M$ and $M^{x+}$ denote the foreign metals and metal ions, and $x$ represents the valence states of the metal ions in the salt solution. Alloying can be achieved at low temperatures (less than 80 °C), or, even at room temperature. In this scenario, the role of temperature is not to mix different elements but to initiate the reduction reaction of metal ions. Considering the reduced metal atoms are distributed on the liquid metal surfaces (at the liquid-liquid interfaces) and the inherent solubility of most metals in Ga-based liquid metal, mixing may occur spontaneously, and a comprehensive library of high-entropy alloys (HEAs) encompassing many elements depicted in the periodic table of elements can be synthesized.

To validate this hypothesis, we selected a thermodynamically immiscible combination of five elements (Pt, Pb, Pd, Au, Cu) (*21*) as representatives to investigate their alloying process with Ga (**Table S1-2**). Before the reaction, the liquid Ga nanoparticles were covered by a uniform layer of gallium oxide due to spontaneous oxidation in the air (see **fig. S1**). Following the reactions, we observed the successful incorporation of all five elements into Ga, resulting in a homogeneous alloy (see **fig. S4**). Moreover, septenary (see **fig. S5**), octonary (see **fig. S6**), and denary (see **fig. S7**) HEA-NPs were successfully synthesized. In **Fig. 1B**, we demonstrate the synthesis of HEAs comprising over 20 elements, each possessing distinct preferred crystal structures, melting points, and atomic radii, thus showcasing the capability to produce diverse HEA-NPs across a wide elemental range. Through high-resolution transmission electron microscopy (HRTEM) and energy-dispersive X-ray spectroscopy (EDX), we have confirmed the compositional homogeneity within the NPs (no phase or elemental segregation) (**Fig. 1B** and **figs. S8, S9**). Remarkably, the formation of HEAs with thermodynamically immiscible elements was accomplished through isothermal solidification to stabilize the liquid alloy state rather than by rapid quenching. Furthermore, the synthesized HEA-NPs are stable at room temperature, as evidenced by the preserved features (size, composition, morphology, and crystallinity) over a storage period of 8 months under room temperature in air (**fig. S10**).

In addition to the mixing of diverse elements, the morphology and crystallinity of these HEAs are varied (**Fig. 1C&D**). By adjusting interfacial reaction kinetics to coordinate the entry and diffusion of metal species into liquid metals, we can synthesize a spectrum of structures spanning from 0D to 3D, including solid, porous, hollow spheres, nanoflowers, nano dendrites, 2D porous nanosheets, as well as 3D networks and hierarchical structures (**Fig. 1C**). Accompanying the morphological changes, the crystallinity can also be modified, enabling the synthesis of HEAs with single-crystalline, polycrystalline, mesocrystalline, and amorphous structures depending on the rate of isothermal solidification (**Fig. 1D**). Since stabilization of the mixing state is induced by composition change that triggers sudden solidification rather than quenching, the crystallinity and morphology of HEAs are controlled by the interfacial reaction kinetics. This fact highlights the different mechanism underlining how various crystalline structures arise in formation of HEAs.

**Controlling crystallinity of HEA-NPs**

Unlike the thermal "shock" methods, solidification in this work is triggered by a composition change in the liquid metal alloy. This suggests that all the reaction parameters that impact how the foreign atoms enter the liquid-liquid interface and their diffusion within the liquid metal alloy can be adjusted to fine-tune the structure as well as crystallinity of the resulting HEAs (**Fig. 1C**). Two notable reaction parameters are temperature and metal salt concentration. We have explored the impact of the rate of composition change on the crystallinity of HEA-NPs (e.g., Ga/Pt/Pd/Au/Cu/Pb) by adjusting the reaction temperature and salt concentration.

This HEA (Ga/Pt/Pd/Au/Cu/Pb) consists of thermodynamically immiscible senary elemental compositions (*4, 21*) (see **Table S1**). Due to variations in atomic size, electronegativity, and lattice parameters of different metals, these metal combinations tend to phase-separate at room temperature (**fig. S11-14**). However, by slightly increasing the reaction temperature to improve the solidification rate, HEA-NPs with homogeneously mixed elements are achieved, which shows a unique combination of crystallinity and morphology (**Fig. 2** and **figs. S15-18**).

At 40 °C, HEA-NPs of single-crystal spheres are generated (**Fig. 2A & B**). HRTEM images and elemental maps (**Fig. 2B**, **fig. S15**) demonstrate that each element is evenly distributed within the particle and forms a single crystal solid solution. In this case, metal atoms, generated through a gentle reaction, dissolve in liquid gallium slowly and mix up thoroughly. When the melting point of the liquid alloy (dependent on its elemental composition) drops below 40°C, the liquid alloy solidifies, forming perfect single crystalline structures.

At higher temperatures (e.g., 60 °C), there is a corresponding increase in the formation, dissolution, and diffusion rates of metal atoms (*22, 23*). Moreover, since $H^+$ ions can react with Ga to generate hydrogen (*24*), the impact of hydrogen gas generation is more pronounced at higher temperatures. Due to the rapid influx of foreign atoms into the liquid metal, the solidification process occurs more quickly, and the hydrogen produced may have no time to escape from the liquid metal before its solidification. Consequently, the trapping of hydrogen within the metal body leads to the formation of a porous structure (**Fig. 2C**, **fig. S16**). Low-magnification TEM and HRTEM images (**Fig. 2C** and **fig. S19**) indicate that the nanoparticles are mesocrystalline and the images show the unique architecture: the large porous spherical particles (100-300 nm) composed of numerous smaller nanoclusters (2-5 nm). Within the mesocrystalline matrix, we deduce that the local lattice distortions and bending arise from stress induced by gas generation and the formidable surface tension of the liquid metal preceding solidification. Notably, the mesocrystalline structure reflects that the dissolved metal atoms can rapidly diffuse and homogeneously mix within the liquid metal (**Fig. 2C**), warranting a single nucleation site and uniform crystal growth.

We further increased the temperature to 80 °C, which resulted in polycrystalline HEA-NPs with a flower-like morphology (**Fig. 2C**, **fig. S17**). In this scenario, the reaction kinetics are even faster. The fast reaction kinetics yield more hydrogen, leading to a more diverse morphology and faster solidification rates. The hydrogen gas escaping from the nanoparticle may contribute to the flower-like morphology with branched structures, and it may also enhance the mixing of metal elements by generating solid-gas interfaces. The overall increased metal ion reduction reactions and metal mixing may promptly increase the alloy composition and, thus the solidification rate. The solidification of the liquid alloy with distinct localized regions (e.g., branches) creates multiple crystal nucleation sites leading to polycrystalline structures (**Fig. 2D**).

Additionally, we increase the concentration of metal in the precursor solution by two times at 40 °C while maintaining other conditions the same (**Fig. 2E**, **fig. S18**). This adjustment boosts the influx of foreign metals at the liquid-liquid interface while preserving their diffusion rate. Consequently, foreign atoms enter the liquid metal and accumulate rapidly, thus promoting solidification. In this case, spherical HEA-NPs with an amorphous structure are obtained (**Fig. 2E**, **fig. S20**). The solid spherical shape indicates that negligible hydrogen gas has entered into the particle during the reaction at 40 °C, akin to the first control experiment. The uniform distribution of metal elements shows that the diffusion rate of metal atoms is comparable to the first control experiment (**Fig. 2B**), and the metal elements are fully mixed. Hence, we deduce that the amorphous structure arises from the swift dissolution of foreign metal atoms, followed by the prompt solidification of the liquid alloy. The faster solidification may stem from a supercooled state within the liquid alloy (*25, 26*). For example, as external metal atoms enter into the liquid metal, the melting point of the liquid alloy gradually decreases below its temperature and achieves a supercooled state by maintaining its liquid form. Once solidification is triggered, it can be completed quickly, yielding an amorphous solid.

It is noted that accompanied by the crystallinity control, HEA-NPs form solid spheres, mesoporous spheres, and flower-like structures. For all these experiments, only liquid Ga metal spheres were used as the precursors.

**Diverse morphologies of HEAs**

Extending from the above HEA-NPs forming different morphology using Ga metal spheres as precursor (**Fig. 2**), we can fabricate HEAs with more diversified morphologies by modifying Ga liquid metal precursor and controlling the reaction conditions. The resulting HEAs can be broadly categorized into 0D, 2D, and 3D morphologies (**Fig. 3**).

HEA-NPs with 0D morphology (including solid, porous, flower-like nanoparticles) are typically obtained by utilizing spherical liquid metal precursors (**Fig. 3A (I, II, IV, & VI)**, Methods). At room temperature (**Fig. 3A (I)**, Methods), when liquid Ga metal is used as precursor (marked by red outline), the resulting HEA-NPs tend to form phase-separated structures resembling egg-yolk and core-shell configurations (**figs. S11-14**). When temperature is increased to 40 °C, HEA-NPs with solid sphere shapes can be attained (marked by yellow outline) in **Fig. 2B** and **Fig. 3A (II)**. Further increasing the temperature to 60°C and 80°C, HEA-NPS with porous spherical, flower-like, dendritic, and snowflake-like morphologies can be obtained (**Fig. 3A (IV & VI)**, **Fig. 2C&D**). This implies that raising the reaction temperature may enhance the uniform distribution of elements, formation of porous structures, and more diversified morphology in high-entropy alloys.

Interestingly, by using liquid Ga alloy instead of pure Ga precursor, we can obtain HEA-NPs with porous structure (highlighted by yellow outline in **Fig. 3A (I &II), fig. S21**) at 40 °C and even room temperature (**fig. S22**). Employing a liquid alloy precursor with a lower melting point yields similar effects as increasing the reaction temperature. This may be attributed to the lower melting point of the liquid gallium alloy, which facilitates atom transport within it, further promoting the reduction reaction at the interface and mixing of foreign atoms. At elevated temperatures (e.g., 60 °C or above), the impact of the liquid metal precursor is not notably pronounced, and both liquid gallium and liquid gallium alloys can yield porous or more diversified structures (**Fig. 3A (IV & VI)**, **Fig. 2C&D, figs. S23&24**).

Without the addition of HCl to the metal salt solution, the formation of the HEA phase at low temperatures (e.g., room temperature, and 40 °C) is hindered, as the oxide on the surface of the liquid metal blocks the interface reaction (*27*). However, increasing the temperature can resolve this issue, allowing for the formation of HEA with various morphologies tracing the original shape of the liquid metal precursor (**Fig. 3A (III & V)**). For instance, ellipsoidal HEA-NPs are generated when spherical liquid metal precursors are employed (marked by indigo outline in **Fig. 3A (III)**), HEAs with irregular shapes are obtained as irregular metal precursors are used (marked by green outline in **Fig. 3A (III)**), and 2D film metal precursors are utilized to craft 2D nanosheets, 2D networks, 3D hierarchical structures, and some other intricate structures (marked by white outline in **Fig. 3A (III & V)**). The reaction between the oxide shell of the liquid metal and water at elevated temperatures, wherein the disruption of the oxide shell creates an opportunity for the reduction reaction at the interface of liquid metal-liquid salt solution, facilitates the formation of HEAs.

**Figs. 3B-D** depict the 0D, 2D, and 3D morphologies of the high entropy alloy as representative examples respectively. Ellipsoid-shaped senary HEA-NPs (**Fig. 3B**, **fig. S25**) with a single crystal structure were synthesized at 60 °C without adding HCl to the metal salt solution. The detailed formation mechanism will be discussed in **Fig. 4**.

**Fig. 3C** and **fig. S26** show the uniform elemental distribution and mesocrystal structure of 2D HEA nanosheets (8: GaCuPbZnAuFeCoAl) at 60 °C. In formation of this morphology, the spontaneous formed oxide layer on the film surface plays a crucial role in preserving the 2D shape. To maintain the 2D surface morphology, we carefully control the reaction kinetics via adjusting the temperature. For example, we ensure the sufficient reaction kinetics to trigger the reaction between Ga and water to form mesocrystal structure. The formation mechanism closely resembles that of porous mesocrystal spheres (**Fig. 2C**). However, the increase of reaction kinetics (e.g., 60 °C) limits the crystallization with sufficient time and the expulsion of $H_2$ bubbles, and thus introduces local voids and strains, resulting in lattice bending and distortion.

At a higher temperature (80 °C), the HEAs show distinct hierarchical structure morphology with an amorphous structure (**fig. S27**). As shown in **Fig. 3D**, the high-entropy alloys are divided into numerous domains resembling "grape bunches". We speculate that this phenomenon derives from the expansion of the liquid metal at high temperatures, leading to the breakdown of the oxide layer into smaller domains. The rapid reaction bypasses crystallization, resulting in direct solidification into an amorphous structure. Furthermore, the structure bears a striking resemblance to deliberately assembled 3D architectures. The basic units are flower-like clusters, which are then organized together to form one-dimensional branches, and the branches are then brought together to form a "grape bunch" (**fig. S28**). In addition to these morphologies, we have also observed some less frequent but more unique morphologies, likely induced by their local environment (**Fig. 3A (V)**), such as feather-like, chrysanthemum flowers, chiral "L" shapes, etc.

**In-situ observation of the hydrogen-gas assisted mixing and isothermal solidification**
We use ellipsoidal single crystal HEA-NPs (5: GaSnInZnCu) as a model system for our in-situ liquid phase TEM experiments, aiming to verify our speculations regarding the alloying mechanisms. To simplify the material system, we employ liquid gallium-based alloy (e.g., GISZ, containing Ga, In, Sn, Zn) as the precursor and introduce only Cu element into the liquid metal to elucidate the alloying mechanism (**Movie S1**). In the initial state (**Fig. 4A**), the liquid metal nanoparticle has a $Ga_2O_3$ shell, which transforms into GaOOH by hydrolysis in the aqueous solution (*28*). Upon heating to 60 °C, the GaOOH shell reacts with the underlayer of Ga generating H$_2$, a process indicated by the emergence of small bubbles at the GaOOH-Ga interface (marked by yellow arrows in **Fig. 4B**). The reaction equations are:

$$Ga_2O_3 + H_2O \rightarrow 2GaOOH \quad (2)$$
$$6GaOOH + 2Ga \rightarrow 4Ga_2O_3 + 3H_2 \uparrow \quad (3)$$

Meanwhile, the Ga is replaced by Cu$^{2+}$ ions from the solution, and the generated Cu metal atoms are incorporated into the liquid metal through the broken part of the GaOOH shell. The small gas bubbles merge to form a larger bubble while the liquid metal moves along the inner wall of the shell (pointed by indigo arrows, **fig. S29**) resulting in uniform mixtures of metal elements (30-58 s). This hydrogen gas-assisted rapid stirring unavoidably enhances metal mixing at the nanoscale (**Movie S2 and fig. S30**), which is distinct compared with previously reported alloying methods. As the reaction proceeds, hydrogen gas consistently emerges, propelling the bubble's expansion (**Fig. 4A&B**). Simultaneously, Ga in the liquid metal is consumed, while copper atoms are continuously entering into the liquid metal (**Fig. 4C**). After 121 s, the fluidity of the liquid metal alloy swiftly declines, transitioning into a crystalline state (**Fig. 4D**). The temperature maintains at 60 °C throughout the alloying and solidification processes. Thus, the solidification can be attributed to an elevation of the melting point induced by the changes in alloy composition. The high-resolution movie also shows that the melting and solidification of metal alloys are related to composition change during the fusion of different domains (**Movie S3 and fig. S31**). After crystallization, the bubbles continue to expand for a brief period before contracting, while the liquid metal undergoes slight reshaping before it is finally stabilized (**Fig. 4C**). HRTEM image (**Fig. 4D**) and EDX characterization (**Fig. 4E**, **fig. S32**) unequivocally demonstrate the homogeneous distribution of metal elements within the resulting alloy, along with its single-crystal structure, thereby affirming it as a HEA-NP. This marks the pioneering direct observation and utilization of spontaneous changes of alloy composition to kinetically trap the High Entropy states during reactions.

**Conclusion**

We have developed a general route for the synthesis of HEAs with controlled composition, crystallinity, and morphology under mild conditions via liquid-liquid interface engineering (**Table S3**). By limiting the reduction reaction from bulk liquid (3D) to liquid-liquid interfaces (2D), this approach significantly decreases the alloying temperature and improves the controllability of the reaction kinetics. It eliminates the trade-off between alloy composition and morphology diversity, thus broadening the scope of high-entropy alloy synthesis and discovery, and enabling formation of previously unattainable HEAs with rich structural configurations. The underlying mechanisms of isothermal solidification and the hydrogen-enhanced alloying process highlight an unexplored pathway to kinetically trap the high entropy state, independent of quenching. Through liquid-liquid interface engineering, it enables unhindered solidification, resulting in the formation of intricate solidification patterns. Further exploration of the diverse HEAs has the potential to expand their applications across broader fields.

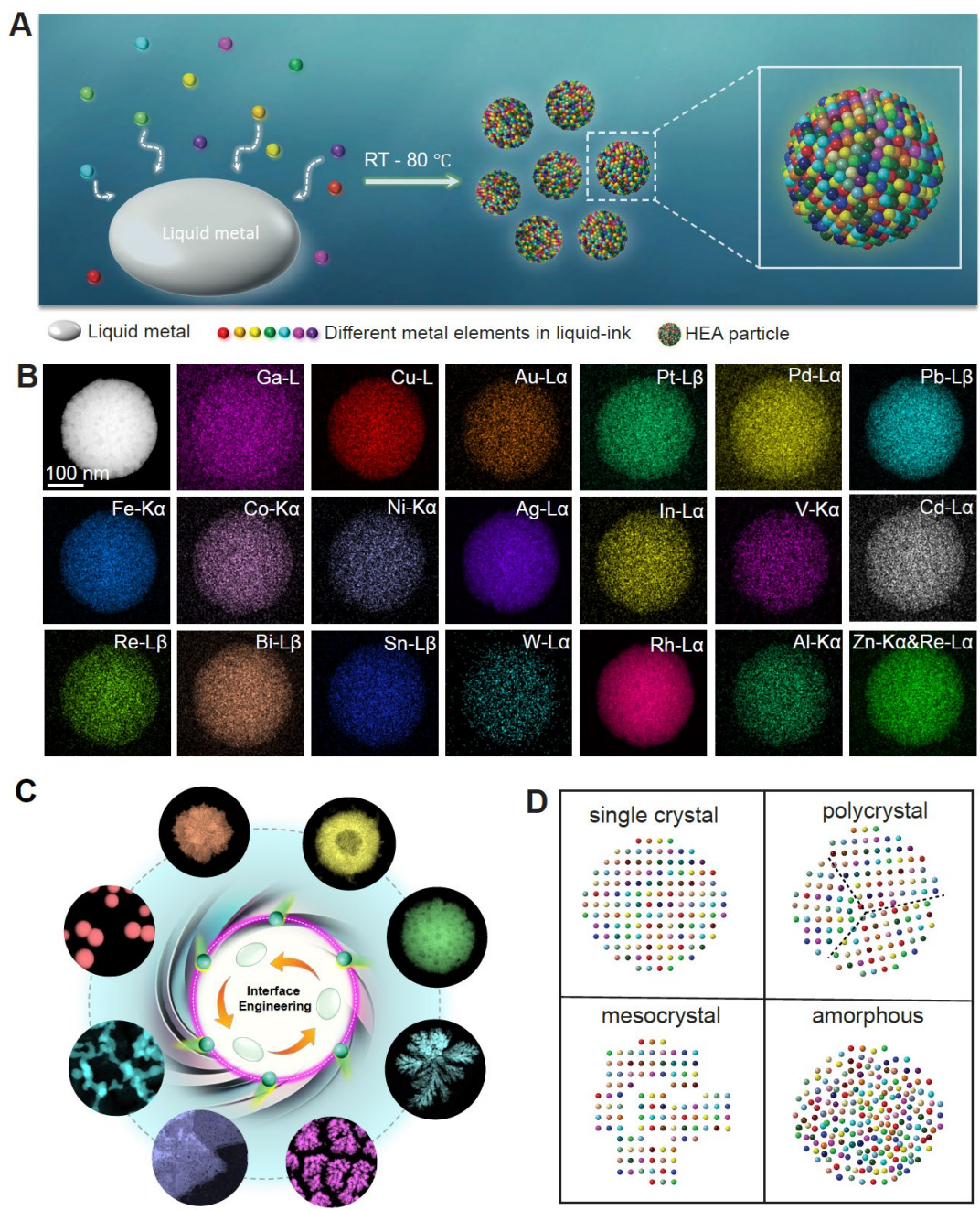

**Fig. 1. Engineering of liquid-liquid interfaces for synthesis of HEA-NPs with controlled crystallinity and morphology.** (**A**) Schematics of HEA-NPs Preparation: Ga-based liquid metal nanoparticles or films are deposited onto a substrate. They are then immersed in a prepared metal salt solution to generate HEA-NPs under mild condition (e.g., RT to 80 °C). (**B**) Elemental maps show HEA-NPs can incorporate as many as 20 dissimilar elements. (**C**) Through liquid-liquid interface engineering, HEA-NPs with diverse shapes and morphologies can be attained by controlling the dissolving and mixing of foreign atoms within the liquid metal. (**D**) Examples of controllable crystallinity (single crystal, mesocrystalline, polycrystalline, and amorphous structures) of synthesized HEA-NPs.

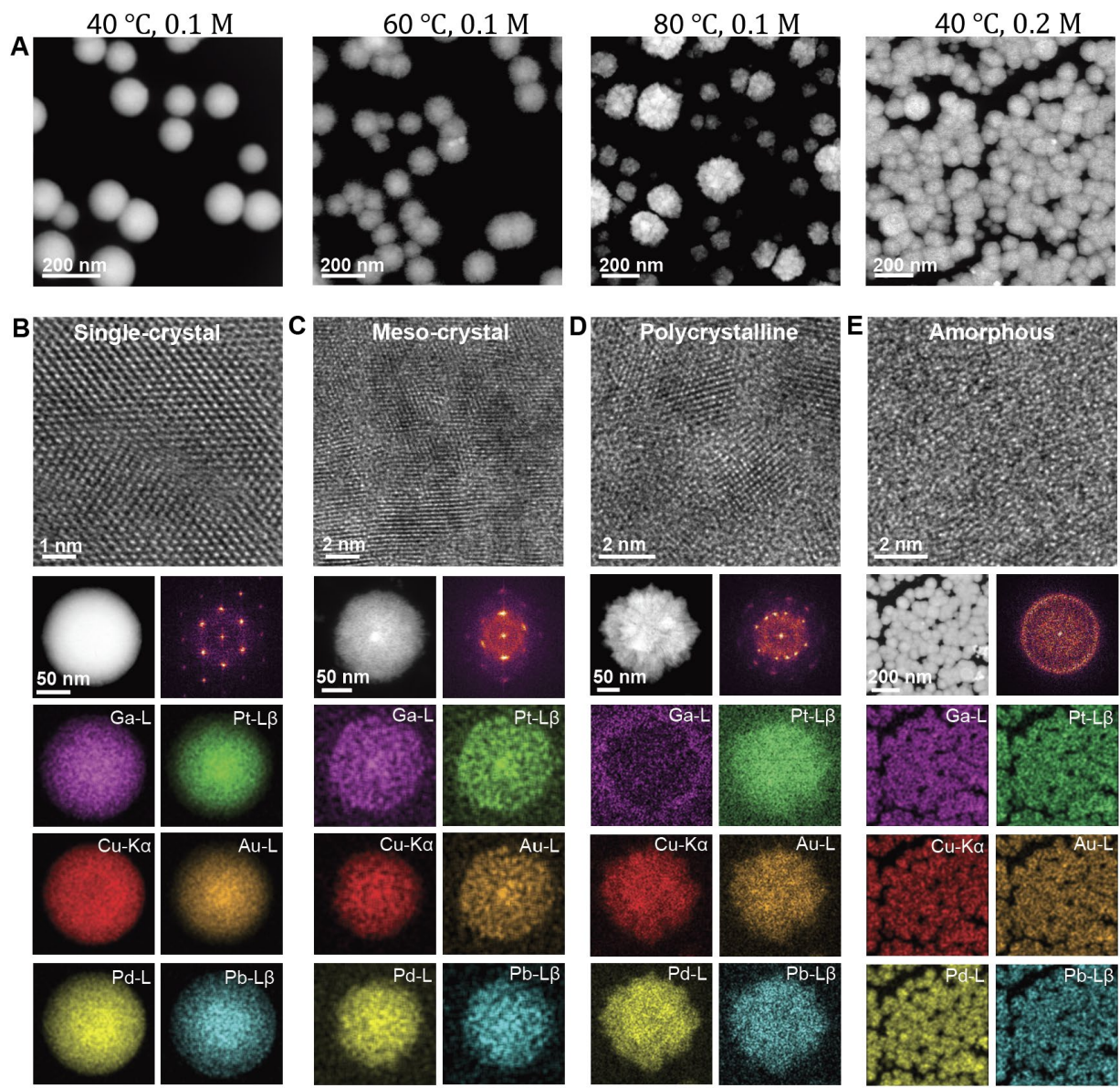

**Fig. 2. Crystallinity control of HEA-NPs.** (**A**) STEM images of synthesized HEA-NPs (GaPtPdPbAuCu) obtained at various temperatures and concentrations of metal ion precursors: 40 °C, 0.1M; 60 °C, 0.1M; 80 °C, 0.1 M; and 40 °C, 0.2 M. (**B-E**) High-resolution TEM images, along with corresponding FFT patterns and EDX mappings, depict representative HEA-NPs synthesized under the following conditions: (B) 40 °C with 0.1 M metal salt concentration, (C) 60 °C with 0.1 M metal salt concentration, (D) 80 °C with 0.1 M metal salt concentration, and (E) 40 °C with 0.2 M metal salt concentration.

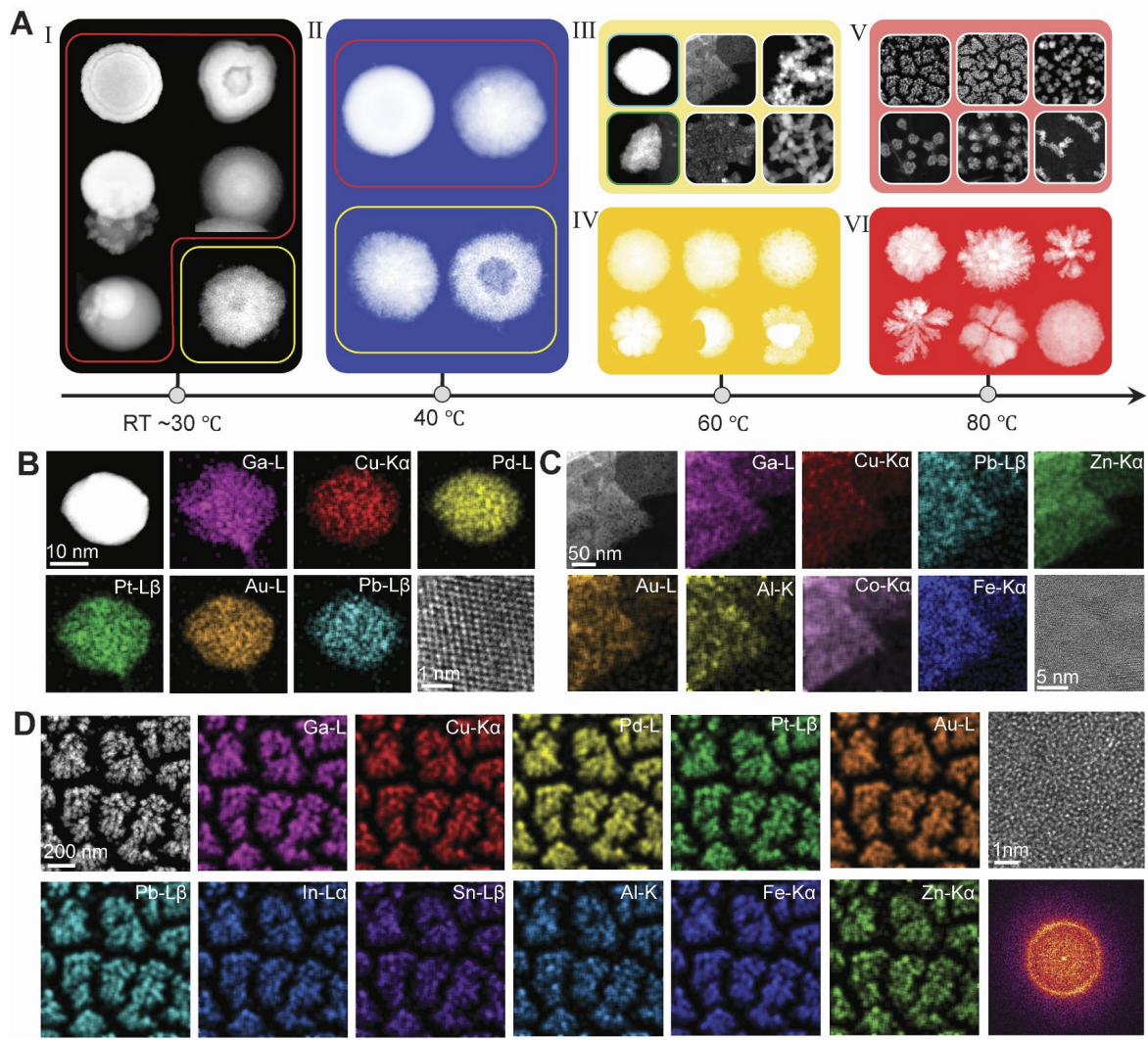

**Fig. 3. Morphology diagram of HEA nanomaterials.** (**A**) STEM images show the diverse morphology of HEAs. **I**: To synthesize nanoparticles in panel (I), we utilize sphere-shaped liquid gallium (marked with red outline) or liquid gallium alloy (marked with yellow outline) as precursors. Additionally, we introduce HCl into the metal salt solution to eliminate the surface oxide of the Ga and Ga alloy particles. **II**, **IV**, **VI**: The synthesis conditions for the particles depicted in panel (II), (IV), and (VI) mirror those in panel (I), but with elevated reaction temperatures of 40 °C, 60 °C, and 80 °C. **III**: we employ spherical shape (marked with indigo outline), irregular shape (marked with green outline), and 2D-film (marked with white outline) liquid Ga/Ga alloy precursors, without the addition of HCl in the metal salt solution, and the reaction temperature is 60 °C. **V**: For HEA synthesis, we use 2D-film liquid Ga/Ga alloy precursors and the reaction temperature is 80 °C. (**B**) STEM elemental maps and HRTEM image of a HEA-NP (6: GaCuPdPtAuPb) representing 0-D nanostructures. (**C**) STEM HAADF image, EDX elemental maps, and HRTEM image of a HEA nanosheet (8: GaCuPbZnAuFeCoAl) representing 2D nanostructures. (**D**) STEM elemental maps and HRTEM image of HEA "grape bunch" structure (11: AlFeCuPtZnPbInSnPdAuGa) representing 3D hierarchical nanostructures.

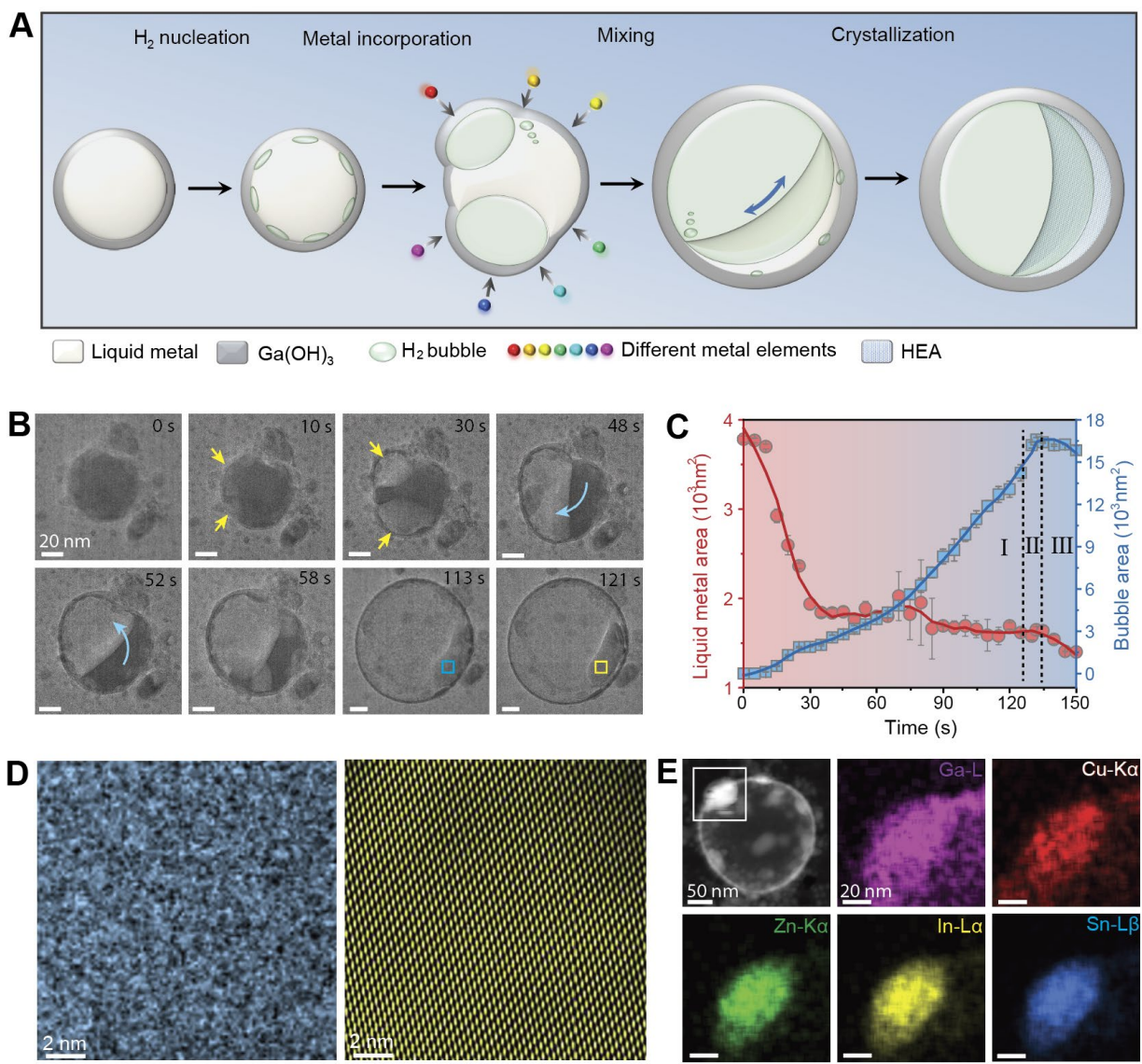

**Fig. 4. The mixing and solidification mechanisms during HEA-NPs formation revealed through in-situ liquid phase TEM.** (**A**) Schematic illustrates the formation process of HEA-NPs, including $H_2$ nucleation, metal incorporation, mixing, and crystallization. (**B**) Sequential in-situ TEM images document the formation details of a HEA-NP (GaInSnZnCu) within a liquid environment at 60 °C. The yellow arrows highlight the nucleation and growth of $H_2$ bubbles, while the indigo arrow indicates the direction of movement of the liquid metal alloy. (**C**) The measured projected areas of the metal alloy and bubble as a function of time. I, II, and III refer to the three intervals of the dynamic process. Data are mean ± standard deviation (s.d.). The data presented are the mean values based on 3 independent measurements; error bars correspond to the s.d.; n=3, 3 independent measurements. (**D**) High-resolution TEM images depicting the structural feature of same local area before and after the solidification of the liquid metal alloy, corresponding to the blue and yellow square areas in **A**. (**E**) EDX mapping shows the elemental distribution of the generated HEA-NP.

**Acknowledgments:**

**Funding:** This work was supported primarily by the U.S. Department of Energy, Office of Science, Office of Basic Energy Sciences (BES), Materials Sciences and Engineering Division under Contract No. DE-AC02-05-CH11231 within the in-situ TEM program (KC22ZH). Part of QZ's effort was supported by the DOE BES funded KCD2S2 program. Work at the Molecular Foundry of Lawrence Berkeley National Laboratory (LBNL) was supported by the U.S. Department of Energy under Contract No. DE-AC02-05CH11231. We thank Diyi Cheng for his help with the schematic drawing of HEA-NP scaling-up synthesis.

**Author contributions:**

Conceptualization: Q.Z., H.Z.

Methodology: Q.Z., Y.C.

Visualization: Q.Z., K.C.B.

Investigation: Q.Z., Y.C., H.Z.

Writing – original draft: Q.Z., Y.C.

Writing – review & editing: Q.Z., K.C.B., H.Z.

Funding acquisition, project administration and supervision: H.Z.

**Competing interests:** Authors declare that they have no competing interests.

**Data and materials availability:** All data needed to evaluate the conclusions of the paper are present in the paper and/or the Supplementary Materials.

**Supplementary Materials**

Materials and Methods

Supplementary Text

Figs. S1 to S32

Tables S1 to S3

References (*29–41*)

Movies S1 to S3